\documentclass[notitlepage,aps,prl,twocolumn,superscriptaddress,amsmath,showpacs,tightenlines,longbibliography,10pt,footnote]{revtex4-1}

\usepackage[]{lineno}
\usepackage{amssymb}
\usepackage{amsmath}
\usepackage{dcolumn}
\usepackage{graphicx}
\usepackage{mathrsfs}
\usepackage{subfigure}
\usepackage{booktabs}
\usepackage{color}
\usepackage{docmute}
\usepackage{hyphenat}
\usepackage{CJKutf8}
\usepackage{multirow}
\usepackage{array}
\usepackage{threeparttable}
\usepackage[mathscr]{euscript}

\usepackage{graphicx}
\usepackage{subfigure}
\usepackage{tikz}
\usetikzlibrary{positioning}
\usepackage{diagbox}

\usepackage{ulem}%删除线
\newcommand{\bea}{\begin{eqnarray}}
\newcommand{\eea}{\end{eqnarray}}
\newcommand{\beq}{\begin{equation}}
\newcommand{\eeq}{\end{equation}}
\newcommand{\nn}{\nonumber}
\def\/{\over}

\usepackage{url}
\usepackage[colorlinks]{hyperref}
\hypersetup{%
	plainpages=true,
	breaklinks=true,%not default in dvips mode, so we must specify
	hypertexnames=false,%not ideal, but needed when pagenums duplicate (`i' vs. `1')
	pageanchor=true,
	bookmarksnumbered=true,
	colorlinks=true,
	linkcolor={blue},
	citecolor={red},
	urlcolor={blue},
	anchorcolor={black}
}

\begin{document}
\title{Extensive Manipulation of Transition Rates and Substantial Population Inversion of Rotating Atoms Inside a Cavity}
\author{Yan Peng}
\thanks{These authors contributed equally to this work.}
\affiliation{Department of Physics, Key Laboratory of Low Dimensional Quantum Structures and Quantum Control of\\ Ministry of Education, and Hunan Research Center of the Basic Discipline for Quantum Effects and Quantum Technologies, Hunan Normal University, Changsha, Hunan 410081, China}

\author{Yuebing Zhou}
\thanks{These authors contributed equally to this work.}
\affiliation{Department of Physics, Key Laboratory of Low Dimensional Quantum Structures and Quantum Control of\\ Ministry of Education, and Hunan Research Center of the Basic Discipline for Quantum Effects and Quantum Technologies, Hunan Normal University, Changsha, Hunan 410081, China}
\affiliation{Department of Physics, Huaihua University, Huaihua, Hunan 418008, China}
\author{Jiawei Hu}
\email[Corresponding author: ]{jwhu@hunnu.edu.cn}
\affiliation{Department of Physics, Key Laboratory of Low Dimensional Quantum Structures and Quantum Control of\\ Ministry of Education, and Hunan Research Center of the Basic Discipline for Quantum Effects and Quantum Technologies, Hunan Normal University, Changsha, Hunan 410081, China}
\author{Hongwei Yu}
\email[Corresponding author: ]{hwyu@hunnu.edu.cn}
\affiliation{Department of Physics, Key Laboratory of Low Dimensional Quantum Structures and Quantum Control of\\ Ministry of Education, and Hunan Research Center of the Basic Discipline for Quantum Effects and Quantum Technologies, Hunan Normal University, Changsha, Hunan 410081, China}

\begin{abstract}
We investigate the transition rates of a centripetally accelerated atom inside a high-quality cavity and show that they can be extensively tuned by adjusting the cavity resonance and the rotation frequency. Crucially, while inertial atoms cannot be excited in vacuum, rotation induces spontaneous excitation via the circular Unruh effect, with the cavity serving only as an amplifier. Using experimentally feasible parameters, we demonstrate that, in one scenario, the excitation rate can reach $\sim 10^7~\mathrm{s}^{-1}$ while emission remains negligible, enabling substantial population inversion.  In another scenario, both excitation and emission can simultaneously attain $\sim 10^7~\mathrm{s}^{-1}$, corresponding to millions of transitions per second for a single atom. These findings highlight a powerful method for manipulating atomic transition rates  for quantum applications and open a promising route toward experimental verification of the circular Unruh effect with state-of-the-art quantum technologies.

\end{abstract}

\maketitle

\emph{Introduction}.---The study of atomic transitions has long been central to our understanding of the interaction between matter and electromagnetic fields within the framework of quantum electrodynamics. In free space, an atom in its excited state can spontaneously emit a photon and transition to a lower energy state, a process known as spontaneous emission. 
 Spontaneous emission in vacuum is governed by the interaction between the atom and the fluctuating electromagnetic field. Consequently, any modification of these vacuum fluctuations leads to changes in the atom's radiative properties.

One particularly intriguing scenario arises when atoms are placed in a resonant cavity. 
The cavity modifies the density of states available for the fluctuating electromagnetic field, significantly enhancing the spontaneous emission rate in a phenomenon known as the Purcell effect~\cite{Purcell46}, which has been experimentally observed~\cite{Purcell2, Purcell3} and has played a pivotal role in advancing our understanding of how confined fluctuating electromagnetic fields influence atomic behavior.

Vacuum fluctuations of the electromagnetic field can be profoundly altered not only by confinement to bounded spaces, such as cavities, but also by the noninertial motion of atoms. Atoms in such motion interact with these modified vacuum fluctuations in unique ways, leading to phenomena that have no counterpart in inertial motion. For instance, an atom undergoing uniform linear acceleration perceives the vacuum, as seen by an inertial observer, as a thermal bath. This leads to the atom spontaneously transitioning to a higher energy level, a phenomenon known as the Unruh effect~\cite{Davies75,Unruh76,Fulling73}. Additionally, the emission rate of an accelerated atom is also enhanced compared to that of an inertial atom, mimicking the behavior of an atom in a thermal environment. However, the excitation rate typically remains much lower than the emission rate, except under extremely high accelerations where it can approach the emission rate. Therefore, while the Unruh effect is theoretically compelling, it has remained elusive in direct experimental observation due to the extreme accelerations required.

At this point, it is important to note that the Unruh effect is significant not only for its intrinsic properties but also for its deep connection to Hawking radiation from black holes~\cite{Hawking74,Hawking75}. Both phenomena are linked through the equivalence principle, which draws a connection between acceleration and gravity.

Similar to the case of linear acceleration, an observer undergoing uniform centripetal acceleration would also perceive the vacuum of an inertial atom as an environment with radiation, although the spectrum of this radiation is nonthermal~\cite{Bell83,Hacyan1986,Bell87,Kim1987,Unruh98,Rosu2005}. Consequently, spontaneous excitation can occur for atoms under centripetal acceleration in a vacuum, a phenomenon that can be regarded as the circular Unruh effect. The emission rate in this scenario can also be enhanced by centripetal acceleration, although the behavior differs significantly from that under uniform linear acceleration~\cite{Letaw1981,Takagi1984,Letaw1980,Rogers88,Davies1996,Lorenci2000,YaoJin2014-1,YaoJin2014-2,Biermann2020,Bunney23,Bunney23-2,yu24}. 

Interestingly, recent studies have shown that when the rotational angular velocity exceeds the transition frequency of the atom, the excitation rate can reach the same order of magnitude as the emission rate, even when the orbital radius and thus the centripetal acceleration are extremely small~\cite{yu24}. 
These findings not only challenge the conventional understanding that a significant Unruh effect requires a large acceleration and reveal that the radiative properties of a rotating atom can be profoundly different from those of inertial atoms, even at vanishingly small centripetal accelerations, but also naturally lead to further questions. Specifically, on one hand, how would the radiative properties of inertial atoms inside a cavity change if they are rotated? 
Can the maximum cavity-amplified emission rate for inertial atoms be further enhanced by centripetal acceleration, as it is in free space? 
On the other hand, how would the radiative properties of rotating atoms in free space change if they are placed inside  a cavity? Can the excitation rate of noninertial atoms be drastically amplified by cavities, similar to how the emission rate is for inertial atoms? 
If possible, can the excitation rate even exceed the emission rate, leading to population inversion? 
These questions lie at the heart of this study, where we explore the interplay between the effects of atomic circular acceleration, a form of noninertial motion, and the confinement of the electromagnetic field within the cavity. Our goal is to control atomic transitions, which are crucial for developing high-brightness single-photon sources for quantum communication, improving  the precision of atomic clocks for quantum metrology,  advancing quantum computing through enhanced qubit readout speeds, and experimentally verifying the circular Unruh effect using state-of-the-art quantum technologies.

\emph{Transition rates of centripetally accelerated atoms in a cavity}.---We consider a two-level atom with a proper transition frequency $\omega_{0}$ coupled to fluctuating vacuum electromagnetic fields within a cavity. Assuming that the atom moves along a circular orbit, its trajectory in the laboratory frame is described by $x(t)=R\cos(\Omega t),~y(t)=R\sin(\Omega t), ~z(t)=0$, 
where $R$ represents the radius of the orbit, $\Omega$ the rotational angular velocity, and $t$ the coordinate time. In the laboratory frame, the Hamiltonian characterizing the dipole interaction between the atom and the electromagnetic field can be expressed as $H_{I}=-\sum_{m=\rho,\phi,z}D_{m} \mathcal{E}_{m}$~\cite{yu24}, 
where $D_{m}$ is the electric dipole moment operator of the atom, and the components of $\mathcal{E}_{m}$ are 
\begin{eqnarray}
\mathcal{E_{\rho}}&=&\Omega R B_{z} +\cos(\Omega t) E_{x} +\sin(\Omega t) E_y,\nonumber\\
\mathcal{E_{\phi}}&=&[\cos(\Omega t)  E_{y} - \sin(\Omega t) E_{x}]/\gamma, \nonumber\\
\mathcal{E}_{z}&=&E_{z}   -\Omega R \sin(\Omega t)B_{y}  -\Omega R \cos(\Omega t)B_{x},
\end{eqnarray}
with $\gamma=\left(1-{\Omega^2 R^2}/{c^2}\right)^{-\frac{1}{2}}$ being the Lorentz factor, $c$ the speed of light, and $E_i$, $B_i$ $(i=x,y,z)$ the components of the electric and magnetic fields in the laboratory coordinate system, respectively. 

Using perturbation theory, the emission rate $\Gamma_{\downarrow}$ and excitation rate $\Gamma_{\uparrow}$ of the atom in the laboratory frame can be derived as follows:
\begin{equation}\label{Gammadownup}
\Gamma_{\downarrow \uparrow}=\sum_{m,n=\rho,\phi,z} \frac{{d}_{m}{d}^*_{n}}{\hbar^2} \int d t_{-}   e ^{\pm i \omega t_{-}}G_{mn}(t_{-})\;,
\end{equation}
where $t_{-}=t_{}-t^{\prime}$, $\hbar$ is the reduced Planck constant, ${d}_{m}=\langle e| D_m|g\rangle$ is the dipole transition matrix element with $|g\rangle$ ($|e\rangle$) being the ground (excited) state of the atom, $G_{mn}(t_{-})=\langle 0|\mathcal{E}_{m}\left(t,\mathbf{x}\right)\mathcal{E}_{n}\left(t',\mathbf{x}'\right)|0\rangle$ are the two-point functions of the electromagnetic fields with $|0\rangle$ being the vacuum state of the electromagnetic fields, and $\omega=\omega_{0}(1-{\Omega^2 R^2}/{c^2})^{\frac{1}{2}}$ is the transition frequency of the atom in the laboratory frame. The $+$ and $-$ signs in the exponential function correspond to the emission rate $\Gamma_{\downarrow}$ and excitation rate $\Gamma_{\uparrow}$, respectively. 
For the explicit forms of the two-point functions 
$\langle 0|\mathcal{E}_{m}\left(t,\mathbf{x}\right)\mathcal{E}_{m}\left(t',\mathbf{x}'\right)|0\rangle$, 
please refer to 
Sec.~I of Supplemental Material. 
These functions are dependent on the explicit form of the density of states within the cavity $\rho(\omega_k)$, where $\omega_k$ denotes the frequency of the field mode. The density of states within a cavity is typically modeled as a Lorentzian distribution given by $\rho(\omega_k)= \frac{1}{\pi}\frac{\omega_c/Q}{(\omega_c/Q)^2 + (\omega_k -\omega_c)^2}$, 
where $Q~(Q\gg1)$ represents the quality factor, and $\omega_c$ denotes the normal mode frequency of the cavity. 
By substituting the explicit forms of the field correlation functions into Eq.~(\ref{Gammadownup}), the leading terms of the emission rate $\Gamma_{\downarrow}$ in the nonrelativistic limit ($v=R\Omega\ll c$) can be calculated as
\begin{eqnarray}\label{eq7}
    &&  \Gamma_{\downarrow}  \approx  \frac{\omega_{c} Q}{6\epsilon_{0}  \hbar V } \bigg \{\frac{2 \omega _0}{Q^2\left(\omega _0-\omega_{c}\right){}^2+\omega_{c}^2}  d_z^2\nonumber\\
    &&\qquad + \left(d_{\rho }^2+d_{\phi }^2\right) \times\left[\frac{\omega _0+\Omega }{Q^2 \left(\omega _0+\Omega -\omega_{c}\right){}^2+\omega_{c}^2} \right.\nonumber\\
    &&\qquad\left.+\frac{ \left(\omega _0-\Omega \right) \Theta \left(\omega _0-\Omega \right)}{Q^2 \left(\omega _0-\Omega -\omega _c\right){}^2+\omega _c^2}\right]\bigg \}+\mathcal{O}\left[({R\Omega}/{c}\right)^2].~~~
\end{eqnarray}
Meanwhile, the excitation rate  $\Gamma_{\uparrow}$ is given by
\begin{eqnarray}
    \label{eq8}
    \Gamma_{\uparrow}\approx\frac{Q \omega_{c} \left(\Omega -\omega _0\right) \left(d_{\rho }^2+d_{\phi }^2\right)}{6 \epsilon_0  \hbar V  \left[Q^2 \left(\Omega-\omega _0 -\omega_{c}\right){}^2+\omega_{c}^2\right]}
    +\mathcal{O}\left[({R\Omega}/{c}\right)^2],~
\end{eqnarray}
for $\Omega > \omega_{0}$. 
Here $\epsilon_0$ is the vacuum permittivity, $V$ represents the volume of the cavity, and $\Theta(x)$ is the Heaviside step function.  
Note that when $\Omega < \omega_{0}$, the leading term of the excitation rate is of the order $({R \Omega}/{c})^2$ or higher, as detailed in 
Sec.~II of Supplemental Material. 
In the following, we discuss explicitly the properties of the transition rates for centripetally accelerated atoms within a cavity, focusing on the interplay between the effects of centripetal acceleration and cavity. It should be noted that the analysis assumes  $Q\gg 1$.

\emph{Rotational effect}---First, we examine the impact of rotation on the radiative properties of atoms by comparing the transition rate of centripetally accelerated atoms inside a cavity with that of inertial atoms in the same environment. 
Since inertial atoms in vacuum cannot be excited regardless of the cavity design,  our focus is on the emission rate. 
In particular, we investigate whether the maximum cavity-amplified emission rate for inertial atoms can be further enhanced by centripetal acceleration.

Rotation-induced suppression at resonance: When the atom is resonant with the cavity, i.e., when $\omega_0=\omega_c$, the emission rate for inertial atoms within the cavity reaches its maximum, 
given by the peak value: 
\begin{equation}\label{g0di}
    \Gamma_{\downarrow}^{\rm in}  = \frac{Q}{3 \epsilon_{0} \hbar V}\left(d_{z}^{2}+d_{\rho}^{2}+d_{\phi}^{2}\right).
\end{equation}
When the atoms are rotated, the emission rate becomes
\begin{align}\label{g1d}
\Gamma_{\downarrow}   \approx &\frac{Q}{3\epsilon_{0} \hbar V }  d_z^2+
\frac{Q\omega_0 [\omega_0+\Omega +(\omega_{0}-\Omega)\, \Theta(\omega_{0}-\Omega) ] }{6\epsilon_{0} \hbar V (Q^2\Omega^2+\omega_{0}^2)} \nonumber\\
&\times(d_\rho^2+d_\phi^2)
+\mathcal{O}[\left({R\Omega}/{c}\right)^2].
\end{align}
As the rotational angular velocity increases from zero, the contribution of the transverse polarization (atomic polarization perpendicular to the rotation axis) to the emission rate is strongly suppressed by rotation. Consequently, the primary contribution to the emission rate arises from the axial polarization (atomic polarization parallel to the rotation axis). According to  Eq.~(\ref{g1d}), rotation always suppresses the emission rate, and the larger the rotational angular velocity, the smaller the emission rate becomes.
This implies that, unlike in free space, the maximum cavity-amplified emission rate for inertial atoms cannot be further enhanced by centripetal acceleration when the atom is resonant with the cavity.

Rotation-induced enhancement off resonance: When the atom is not resonant with the cavity, 
 the emission rate, as described by  Eq.~(\ref{eq7}),  initially increases with the rotational angular velocity. After reaching a peak, it decreases as the rotational speed continues to increase. The peak appears at $\Omega=\omega_c-\omega_0$ when $\omega_0<\omega_c$, and at $\Omega=\omega_0-\omega_c$ when $\omega_0>\omega_c$. In either case, the peak value of the emission rate is 
\begin{equation}\label{g0d1}
\Gamma_{\downarrow}  \approx \frac{Q}{6 \epsilon_{0} \hbar V}\left(d_{\rho}^2+d_{\phi}^2\right)
+\mathcal{O}[\left({R\Omega}/{c}\right)^2],
\end{equation}
where it is assumed that $|\omega_0-\omega_c|\gg \omega_c/Q$.  In this scenario, the leading contribution to the emission rate comes from the transverse polarization. 
For inertial atoms in the same cavity, the emission rate is given by
\begin{equation}
    \Gamma_{\downarrow}^{\rm in} \approx \frac{1}{3\epsilon_{0} \hbar V Q}\frac{\omega _0 \omega _c}{ (\omega _0-\omega _c)^2}\left(d_{z}^{2}+d_{\rho}^{2}+d_{\phi}^{2}\right).
\end{equation} 
Given that $Q\gg1$, the spontaneous emission is significantly enhanced by rotation 
if the atom's transition frequency $\omega_0$ is not close to the cavity's normal mode frequency $\omega_c$ (i.e., $|\omega_0-\omega_c|\gg \omega_c/Q$). 
This significant enhancement occurs despite the emission rate being dominated by the transverse polarization, which is markedly different from the behavior observed in inertial atoms. 
Both   Eq.~(\ref{g1d}) and   Eq.~(\ref{g0d1}) are consistently smaller than  Eq.~(\ref{g0di}), suggesting that the maximum cavity-amplified emission rate for inertial atoms cannot be further enhanced by rotation, regardless of whether the atom is resonant with the cavity.

\emph{Cavity effect}---Next, we investigate how the transitions of centripetally accelerated atoms can be significantly influenced by the presence of the cavity. 
To do this, we compare the peak values of the emission and excitation rates of these atoms when they are inside a cavity with those when they are in free space. 

Significant enhancement of emission rate: We start with the emission rate. 
As shown in Eq.~(\ref{eq7}), when the rotational angular velocity $\Omega$ exceeds the transition frequency of the atom $\omega_0$, i.e., when $\Omega > \omega_0$, the emission rate as a function of the cavity's normal mode frequency exhibits two peaks, located at $\omega_c=\omega_0$ and $\omega_c=\omega_0+\Omega$, respectively. When $\Omega \leq \omega_{0}$, a third peak appears at $\omega_c = \omega_0 - \Omega$. 
According to Eq.~(\ref{g1d}) and Eq.~(\ref{g0d1}), for isotropically polarizable atoms with $d_{z}^{2}=d_{\rho}^{2}=d_{\phi}^{2}\equiv d^2$, the leading terms of the peak emission rates at these points take the same value, which is $\frac{Qd^2}{3\epsilon_{0} \hbar V }$. 
Comparing $\Gamma_{\downarrow}=\frac{Qd^2}{3\epsilon_{0} \hbar V }$ with 
the emission rate of a rotating atom in free space as shown in Eqs.~(15)-(17) of Ref.~\cite{yu24}
reveals that the emission rate of rotating atoms with isotropic polarizability can be enhanced by a factor of $\frac{Q c^3}{V \omega_{0}^{3}}$ when  $\Omega \leq  \omega_{0}$, and by a factor of $\frac{Q c^3}{V (\Omega+\omega_{0})^{3}}$ when  $\Omega > \omega_{0}$, respectively. These enhancement factors can be extremely large, as $Q \gg 1$ and the cavity volume can be very small. As a result, the emission rate for rotating atoms inside the cavity is significantly amplified compared to that in free space~\cite{yu24}.

Significant enhancement of excitation rate: Next, we examine whether the excitation rate of centripetally accelerated atoms can also be significantly amplified by the  presence of the cavity. 
According to Eq.~(\ref{eq8}), when $\Omega > \omega_{0}$, 
the excitation rate peaks when the normal mode frequency of the cavity is tuned to $\omega_{c}=\Omega-\omega_{0}$, with the peak value given by
\begin{eqnarray}\label{eq16}
\Gamma_{\uparrow}  \approx \frac{Q}{6  \epsilon_{0} \hbar V}\left(d_{\rho}^2+d_{\phi}^2\right)
+\mathcal{O}[\left({R\Omega}/{c}\right)^2],
\end{eqnarray}
which coincides with the peak value of the emission rate Eq.~(\ref{g0d1}) at $\omega_{c}=\omega_{0}+\Omega$, indicating that the excitation rate for rotating atoms inside the cavity can be drastically amplified, 
even reaching the maximum cavity-amplified emission rate for rotating atoms. 
For comparison, the excitation rate of a centripetally accelerated atom in free space when  $\Omega > \omega_{0}$ is given by 
Eq.~(17) in Ref.~\cite{yu24}.
Therefore, the maximum excitation rate of a centripetally accelerated atom inside a cavity is approximately $\frac{Q c^3}{V(\Omega-\omega_{0})^3}$ times that in free space. 
When $\frac{\omega_0}{2}<\Omega \leq \omega_{0}$, this enhancement factor becomes $\frac{Q c^3}{V (2 \Omega -\omega _0)^{3}}$, as detailed in 
Sec.~II of Supplemental Material. 
This indicates that the excitation rate can be significantly amplified by a high-$Q$ cavity, regardless of whether the rotational angular velocity exceeds the atom's transition frequency.

Simultaneous enhancement of emission and excitation rates:   Generally, the peaks of the emission and excitation rates do not overlap. However, they can coincide when the rotational angular velocity takes specific values. For   $\Omega>\omega_0$, the peak of the excitation rate at $\omega_{c}=\Omega-\omega_{0}$ coincides with the peak of the emission rate at $\omega_{c}=\omega_{0}$ when $\Omega=2\omega_{0}$. For isotropically polarizable atoms, the leading term of the emission rate Eq.~(\ref{g1d}) is the same as that of the excitation rate Eq.~(\ref{eq16}). This suggests that both the emission and excitation rates can be significantly enhanced simultaneously in a high-$Q$ cavity. 
For the case $\Omega\leq\omega_0$, the emission and excitation rates can also be simultaneously enhanced, while the excitation rate is always smaller than the emission rate (see 
Sec.~II of Supplemental Material).

Population inversion: Now, we further investigate whether the cavity-enhanced excitation rate can surpass the emission rate. As discussed previously, this cannot occur when the peaks of the emission and excitation rates coincide.  Thus, we focus on cases when the rotational angular velocity $\Omega$ is not in the vicinity of $2\omega_{0}$, thereby avoiding overlap between the peaks of the emission and excitation rates. According to Eq.~(\ref{eq7}), when $\Omega$ is not near $2\omega_{0}$, the emission rate for rotating atoms inside a cavity with the normal mode frequency tuned to $\omega_c=\Omega-\omega_0$ is given by
\begin{align}
\label{g3d}
\Gamma_{\downarrow}  \approx &\frac{\Omega ^2-\omega _0^2}{24 \epsilon_{0} \hbar V  Q \omega _0^2} \left[d_{\rho }^2+d_{\phi }^2+\frac{8 \omega _0^3 }{\left(\Omega -2 \omega _0\right){}^2 \left(\Omega +\omega _0\right)}d_z^2 \right]\nn\\
&+\mathcal{O}[\left({R\Omega}/{c}\right)^2],
\end{align}
where it is assumed that the quality factor is very large such that $Q\gg1$ and $Q\gg\Omega/\omega_0$. 
Comparing Eq.~(\ref{eq16}) and Eq.~(\ref{g3d}), the excitation rate is approximately $\frac{Q^2}{(\Omega/\omega_0)^2-1}$ times that of the emission rate when $\Omega \gtrsim 3\omega_0$. 
Remarkably, this suggests that the excitation rate can be significantly greater than the emission rate, given that $Q \gg 1$. Therefore, significant population inversion can be achieved  for centripetally accelerated atoms within  a high-$Q$ cavity through the combined effects of rotation and cavity enhancement:  while static atoms in a cavity show no spontaneous excitation and rotating atoms in free space exhibit excitations that never surpass emissions, their interplay in the cavity enables the higher energy state to become more populated than the lower one.
Population inversion is also achievable when $\Omega\leq\omega_0$ (see 
Sec.~II of Supplemental Material).

Suppression of transition rates: In addition to enhancing transition rates, the cavity can also significantly suppress them. See 
Sec.~III of Supplemental Material for a discussion.

\emph{Discussion}---We have the following comments based on the investigations  presented  above.

It is well known that in a vacuum, an inertial atom in its ground state cannot spontaneously transition to an excited state. However, this is possible for atoms that are uniformly accelerated. 
From an operational perspective, the spontaneous transition of an accelerated atom from the ground state to an excited state in a vacuum is a direct manifestation of the Unruh effect~\cite{Davies75,Unruh76,Fulling73}. The Unruh effect is fundamentally significant but has proven elusive in experimental detection, primarily because the spontaneous excitation is too faint at the accelerations achievable in current experiments. Although it has been shown in Ref.~\cite{yu24} that the excitation rate of a rotating atom in a vacuum can reach the same order of magnitude as the emission rate when the rotational angular velocity exceeds the atomic transition frequency, resulting in  excitation rate $10^{272,878}$ times higher than that of an atom under uniform linear acceleration, the actual excitation rate still remains exceedingly weak, posing significant challenges for laboratory observation.
To illustrate this explicitly, we take 
$d = 10^{-29} \, \text{Cm}$, $V=10^{-14} \, \text{m}^3$, $Q = 10^7$, $R = 50 \, \text{nm}$, $ \Omega=5\, \text{GHz}$, and $\omega_{0} = 10 \, \text{MHz}$, satisfying $R \Omega /c\ll 1$. Note that these parameters are experimentally feasible, and consistent with Ref.~\cite{Lochan20}. 
With these values, the excitation rate in free space is found to be only  $10 ^{-11}\,{\text{s}}^{-1} $, which is far below the threshold for experimental detection. In stark contrast, if the rotating atom is placed inside a cavity with the normal mode frequency tuned to $\omega_c=\Omega-\omega_0$, 
the excitation rate can be amplified to $10 ^{7}~{\text{s}}^{-1}$ according to Eq.~(\ref{eq16}). At this level, the excitation rate becomes readily observable in experiments. This represents an enhancement of 18 orders of magnitude compared to rotating atoms in free space. 
Remarkably, this peak value is independent of the atom's transition frequency and rotational speed, making the choice of parameters less restrictive. The only requirement is the condition $\omega_c=\Omega-\omega_0$. 
Meanwhile, the emission rate in this case is negligible, trailing the excitation rate by 9 orders of magnitude. 
This result highlights the potential for substantial population inversion. 

In addition to measuring the cavity-enhanced spontaneous excitation as previously discussed, the circular Unruh effect can also be verified by detecting cavity-enhanced spontaneous emission. By tuning the normal mode frequency of the cavity to $\omega_c=\Omega+\omega_0$ and using experimentally feasible parameters as before, it is possible to observe emission specifically due to centripetal acceleration, which can reach as high as $10 ^{7}\,{\text{s}}^{-1}$. 
More interestingly, by setting the rotational angular velocity and the cavity's normal mode frequency to $\Omega=2\omega_c=2\omega_0=5 \, \text{GHz}$ while keeping the other parameters unchanged, both the emission and excitation rates can reach as high as $\sim 10^{7} \, \text{s}^{-1}$  simultaneously. 
This configuration can result in millions of transition events per second for a single centripetally accelerated atom, offering a robust method for experimentally verifying the circular Unruh effect. Such a phenomenon is impossible for an inertial atom, even if its emission rate is higher when $\omega_c = \omega_0$, because once it emits a photon and transitions to the ground state, it cannot be excited again. In contrast, the continuous transitions between the ground and excited states of a single centripetally accelerated atom serve as compelling evidence for the circular Unruh effect.

Based on the discussions above, we  argue that manipulating the transition rates and achieving significant population inversion are indeed feasible with current state-of-the-art technologies.  Recently, hyperfast rotation of an optically levitated nanoparticle has been achieved by transferring the spin angular momentum of light to the particle's mechanical angular momentum~\cite{Reimann2018,Li2018,Zhang2021}.  This  progress  opens up the promising possibility of attaching an atom to such a rapidly rotating, optically levitated nanoparticle. Notably, the physical parameters used in our numerical investigations  ($R = 50 \, \text{nm}$ and $ \Omega=5 \, \text{GHz}$) align well with recent experimental developments~\cite{Zhang2021}, suggesting that the proposed experiments could be practically realizable. 

There are recent studies focusing on the radiative properties of accelerated atoms within a cavity~\cite{Scully03,Lochan22,Lochan20,guo24,Arya23,Arya24}. Notably, in a recent work~\cite{Lochan20} where only the emission of excited atoms is investigated, it was demonstrated that the emission rate of a rotating atom due to centripetal acceleration within a cavity can be significantly amplified by the cavity. However, the emission rate reported in Ref.~\cite{Lochan20} is substantially lower---by 14 orders of magnitude---than the rate found in the current study when using the same parameters, necessitating an ensemble of $10^6$ atoms to observe significant transitions. 
This significant discrepancy arises because, in Ref.~\cite{Lochan20}, the atoms are assumed to be polarizable only along the direction orthogonal to the plane of rotation. In contrast, our study assumes isotropic polarizability, which is more common in nature.   In the nonrelativistic limit ($R\Omega/c\ll1$), the leading term for the emission rate when $\omega_c=\Omega+\omega_0$ is given by Eq.~(\ref{g0d1}), where the dominant contribution comes from transverse polarization. This rate is approximately $\sim c^2/(R \Omega)^2~(\gg 1)$ times that from axial polarization, making the axial contribution calculated in Ref.~\cite{Lochan20} negligible.

In practice, the environment is a thermal bath not a perfect vacuum. 
At room temperature ($T = 300\,\mathrm{K}$), we find that the peak excitation rate of centripetally accelerated atoms can reach $10^{11}\;\mathrm{s}^{-1}$ at $\omega_c = \Omega \pm \omega_0$, in stark contrast to only $10^{-3}\;\mathrm{s}^{-1}$ for inertial atoms—an enhancement of 14 orders of magnitude due to centripetal acceleration. Further details are provided in Sec. IV of Supplemental Material. This sharp contrast demonstrates that directly comparing rotating and static cases offers an experimentally practical and robust strategy  to distinguish Unruh-induced excitations from thermal effects, even at room temperature.

Finally, we note that rotation is also linked to another striking phenomenon, superradiance—pioneered  by Zel’dovich for rotating bodies \cite{Zeldovich1971,Zeldovich1972} and recently observed experimentally \cite{NaturePhys2020}—which stands as a milestone in the study of rotation-induced phenomena.  Unlike superradiance, which is a classical wave-scattering process, the circular Unruh effect studied here is a genuinely quantum phenomenon: it concerns spontaneous excitation of a rotating detector by vacuum fluctuations, even in the absence of incident radiation.

\emph{Summary}---We have demonstrated that the interplay between atomic centripetal acceleration and the confinement of the electromagnetic fields within a high-$Q$ cavity can profoundly alter the radiative properties of atoms. 
While the maximum cavity-amplified emission rate for inertial atoms cannot be further enhanced by rotation, the spontaneous emission can still be significantly amplified if the atom is far from resonance with the cavity.
Notably, the transition rates change dramatically when rotating atoms are placed inside the cavity, enabling significant enhancement of either the emission or excitation rates, individually or even simultaneously.

Remarkably, with realistic physical parameters, we show that, in one scenario, the excitation rate can reach values as high as $10^7~{\rm s}^{-1}$, surpassing those in free space by 18 orders of magnitude.  
Meanwhile, the emission rate is negligible compared with the excitation rate, trailing by 9 orders of magnitude, which suggests the potential for significant population inversion in an ensemble of atoms. 
In another scenario, the emission and excitation rates can simultaneously reach magnitudes as high as $10^7~{\rm s}^{-1}$, indicating that millions of transitions per second are possible even for a single atom. 

These results suggest a powerful route to photon-emission enhancement for quantum technologies and a feasible pathway toward experimental verification of the circular Unruh effect.

\emph{Acknowledgments}---This work was supported in part by the NSFC under Grant No. 12075084, No. 12375047, and No. 12575051; the innovative research group of Hunan Province under Grant No. 2024JJ1006; the Hunan Provincial Natural Science Foundation of China under Grant No. 2023JJ40515; and the Scientific Research Program of Education Department of Hunan Province of China under Grant No. 22B0762.

\emph{Data availability}---No data were created or analyzed in this study.

\clearpage
\appendix
\begin{widetext}
\begin{center}
{\bf Supplemental Material for ``Extensive Manipulation of Transition Rates and Substantial Population Inversion of Rotating Atoms Inside a Cavity''}
\end{center}
\end{widetext}

\renewcommand{\theequation}{A\arabic{equation}}
\section*{I. The two-point functions of the fluctuating electromagnetic fields in a cavity}
\label{Appendix A}

The transition rates are dependent on the two-point functions of the electromagnetic fields $G_{mn}(t_{-})=\langle 0|\mathcal{E}_{m}\left(t,\mathbf{x}\right)\mathcal{E}_{n}\left(t',\mathbf{x}'\right)|0\rangle$. For simplicity, if we assume the transition matrix elements ${d}_{m}$ are real, then the cross terms in  
Eq.~(2) of the Letter
vanish, and the diagonal 
components of the two-point correlation functions for the fluctuating electromagnetic fields in a cavity are shown as follows,
\begin{widetext}
\begin{eqnarray}
\langle 0|\mathcal{E}_{\rho}\left(t,\mathbf{x}\right)\mathcal{E}_{\rho}\left(t',\mathbf{x}'\right) |0\rangle
&=& \Omega^2 R^2\langle 0|B_{z}\left(t,\mathbf{x}\right)B_{z}\left(t',\mathbf{x}'\right) |0\rangle+ \Omega R \cos(\Omega t')\langle 0|B_{z}\left(t,\mathbf{x}\right)E_{x}\left(t',\mathbf{x}'\right)|0\rangle\nonumber\\
&&+ \Omega R \sin(\Omega t')\langle 0|B_{z}\left(t,\mathbf{x}\right)E_{y}\left(t',\mathbf{x}'\right) |0\rangle+ \Omega R \cos(\Omega t)\langle 0|E_{x}\left(t,\mathbf{x}\right)B_{z}\left(t',\mathbf{x}'\right) |0\rangle\nonumber\\
&&+ \cos(\Omega t)\cos(\Omega t')\langle 0|E_{x}\left(t,\mathbf{x}\right)E_{x}\left(t',\mathbf{x}'\right)|0\rangle
+ \cos(\Omega t)\sin(\Omega t')\langle 0|E_{x}\left(t,\mathbf{x}\right)E_{y}\left(t',\mathbf{x}'\right) |0\rangle\nonumber\\
&&+ \Omega R \sin(\Omega t)\langle 0|E_{y}\left(t,\mathbf{x}\right)B_{z}\left(t',\mathbf{x}'\right) |0\rangle
+ \sin(\Omega t)\cos(\Omega t')\langle 0|E_{y}\left(t,\mathbf{x}\right)E_{x}\left(t',\mathbf{x}'\right)|0\rangle\nonumber\\
&&+ \sin(\Omega t)\sin(\Omega t')\langle 0|E_{y}\left(t,\mathbf{x}\right)E_{y}\left(t',\mathbf{x}'\right) |0\rangle\;,
\end{eqnarray}
\begin{align}
\langle 0|\mathcal{E}_{\phi}\left(t,\mathbf{x}\right)\mathcal{E}_{\phi}\left(t',\mathbf{x}'\right)|0\rangle
=&\gamma^{-2}\left[\cos(\Omega t)\cos(\Omega t')\langle 0|E_{y}\left(t,\mathbf{x}\right)E_{y}\left(t',\mathbf{x}'\right) |0\rangle
-\cos(\Omega t)\sin(\Omega t')\langle 0|E_{y}\left(t,\mathbf{x}\right)E_{x}\left(t',\mathbf{x}'\right) |0\rangle\right.\nonumber\\
&\left.-\sin(\Omega t)\cos(\Omega t')\langle 0|E_{x}\left(t,\mathbf{x}\right)E_{y}\left(t',\mathbf{x}'\right) |0\rangle
+\sin(\Omega t)\sin(\Omega t')\langle 0|E_{x}\left(t,\mathbf{x}\right)E_{x}\left(t',\mathbf{x}'\right) |0\rangle\right]\;,
\end{align}
\begin{eqnarray}
\langle 0|\mathcal{E}_{z}\left(t,\mathbf{x}\right)\mathcal{E}_{z}\left(t',\mathbf{x}'\right)|0\rangle 
&=& \langle 0|E_{z}\left(t,\mathbf{x}\right)E_{z}\left(t',\mathbf{x}'\right) |0\rangle
- \Omega R\sin(\Omega t')\langle 0|E_{z}\left(t,\mathbf{x}\right)B_{y}\left(t',\mathbf{x}'\right) |0\rangle\nonumber\\
&&- \Omega R\cos(\Omega t')\langle 0|E_{z}\left(t,\mathbf{x}\right)B_{x}\left(t',\mathbf{x}'\right) |0\rangle
- \Omega R\sin(\Omega t)\langle 0|B_{y}\left(t,\mathbf{x}\right)E_{z}\left(t',\mathbf{x}'\right) |0\rangle\nonumber\\
&&+\Omega^2 R^2 \sin(\Omega t)\sin(\Omega t')\langle 0|B_{y}\left(t,\mathbf{x}\right)B_{y}\left(t',\mathbf{x}'\right) |0\rangle\nonumber\\
&&+\Omega^2 R^2 \sin(\Omega t)\cos(\Omega t')\langle 0|B_{y}\left(t,\mathbf{x}\right)B_{x}\left(t',\mathbf{x}'\right) |0\rangle\nonumber\\
&&- \Omega R\cos(\Omega t)\langle 0|B_{x}\left(t,\mathbf{x}\right)E_{z}\left(t',\mathbf{x}'\right) |0\rangle
+\Omega^2 R^2 \cos(\Omega t)\sin(\Omega t')\langle 0|B_{x}\left(t,\mathbf{x}\right)B_{y}\left(t',\mathbf{x}'\right) |0\rangle\nonumber\\
&&+\Omega^2 R^2 \cos(\Omega t)\cos(\Omega t')\langle 0|B_{x}\left(t,\mathbf{x}\right)B_{x}\left(t',\mathbf{x}'\right) |0\rangle\;,
\end{eqnarray}
where
\begin{eqnarray}
&&\left\langle 0\left|E_{l}(t,\mathbf{x}) E_{p}\left(t',\mathbf{x}'\right)\right| 0\right\rangle=\frac{\hbar}{8 \pi \epsilon_{0} V}\int_{0}^{2 \pi} d\varphi \int_{0}^{\pi} \sin\theta d\theta \int_{0}^{\infty} d\omega_{k} \rho(\omega_k) \frac{\omega_k}{2} \left(\delta_{lp}-\frac{k_{l}k_{p}}{\boldsymbol{k}^{2}}\right) e^{-i( \omega_{k} t_{-}-\boldsymbol{k} \cdot \boldsymbol{R})}\label{correlationfunctionEE}\;,\\
&&\left\langle 0\left|B_{l}(t,\mathbf{x}) B_{p}\left(t',\mathbf{x}'\right)\right| 0\right\rangle=\frac{\hbar}{8 \pi\epsilon_{0} V}\int_{0}^{2 \pi} d\varphi \int_{0}^{\pi} \sin\theta d\theta \int_{0}^{\infty} d\omega_{k} \rho(\omega_k) \frac{\omega_k}{2c^2} \left(\delta_{lp}-\frac{k_{l}k_{p}}{\boldsymbol{k}^{2}}\right)  e^{-i( \omega_{k} t_{-}-\boldsymbol{k} \cdot\boldsymbol{R})}\label{correlationfunctionBB}\;,\\
&&\left\langle 0\left|E_{l}(t,\mathbf{x}) B_{p}\left(t',\mathbf{x}'\right)\right| 0\right\rangle=\frac{\hbar}{8 \pi \epsilon_{0} V}\int_{0}^{2 \pi} d\varphi \int_{0}^{\pi} \sin\theta d\theta\int_{0}^{\infty} d\omega_{k} \rho(\omega_k) \frac{\omega_k}{2c} \epsilon_{lpq} \frac{k_{q}}{\lvert \boldsymbol{k}\rvert} e^{-i( \omega_{k} t_{-}-\boldsymbol{k} \cdot\boldsymbol{R})}\label{correlationfunctionEB}\;,\\
&&\left\langle 0\left|B_{l}(t,\mathbf{x}) E_{p}\left(t',\mathbf{x}'\right)\right| 0\right\rangle=\frac{\hbar}{8 \pi \epsilon_{0} V}\int_{0}^{2 \pi} d\varphi \int_{0}^{\pi} \sin\theta d\theta\int_{0}^{\infty} d\omega_{k} \rho(\omega_k) \frac{\omega_k}{2c} \left(-\epsilon_{lpq} \frac{k_{q}}{\lvert \boldsymbol{k}\rvert}\right)  e^{-i( \omega_{k} t_{-}-\boldsymbol{k} \cdot\boldsymbol{R})}\label{correlationfunctionBE}\;.
\end{eqnarray}
\end{widetext}
In the formulae shown above, $l,p,q=x,y,z$, $\hbar$ is the reduced Planck constant, $\epsilon_0$ is the vacuum permittivity, $V$ is the volume of the cavity, $\omega_k$ is the frequency of the field mode, $\boldsymbol{k}$ is the mode vector, $ \epsilon_{lpq}$ is the Levi-Civita symbol, the term $\boldsymbol{R}=\mathbf{x}(t)-\mathbf{x}(t^{\prime})$ denotes a time-dependent displacement vector as the atom undergoes rotation,  
$t_{-}=t_{}-t^{\prime}$, and 
$\rho(\omega_k)$ is the density of states function in the cavity, which is taken as the Lorentzian form in this Letter.

\section*{II. The excitation rate when $\Omega<\omega_0$}\label{Appendix B}

The excitation rate  $\Gamma_{\uparrow}$ is given by 
\begin{equation}\label{eq9}
    \Gamma_{\uparrow}\approx\frac{Q R^2 \omega_{c} \left(2 \Omega -\omega _0\right){}^3 \left(d_{\rho }^2+d_{\phi }^2\right)}{40 c^2  \epsilon_0  \hbar V \left[Q^2 \left(2 \Omega -\omega _0-\omega_{c}\right){}^2+\omega_{c}^2\right]}
    +\mathcal{O}[({R \Omega}/{c})^4]\;,
\end{equation}
for  $\frac{\omega_0}{2} < \Omega \leq \omega_{0}$. When $0<\Omega\leq\frac{\omega_0}{2}$, the leading term of the excitation rate is of order $({R \Omega}/{c})^4$ or higher, and this case is therefore not considered.

The excitation rate Eq.~\eqref{eq9} reaches its peak when the normal mode frequency of the cavity is tuned to $\omega_c={2\Omega-\omega}_{0}$, with the peak value given by the following expression:
\begin{equation}\label{g0u3}
\Gamma_{\uparrow}  \approx \frac{Q R^2 \left(2 \Omega -\omega _0 \right){}^2 }{40 \epsilon_{0} \hbar V c^2 }\left(d_{\rho}^2+d_{\phi}^2\right)+\mathcal{O}[\left({R\Omega}/{c}\right)^4]\;.
\end{equation}
For comparison, the corresponding excitation rate for rotating atoms in free space is
\begin{equation}\label{g0uf}
\Gamma_{\uparrow}^{\rm free}  \approx \frac{R^2 \left(2 \Omega -\omega _0\right){}^5}{40 \pi \epsilon_{0} \hbar c^5} \left(d_{\rho}^2+d_{\phi}^2\right)+\mathcal{O}[\left({R\Omega}/{c}\right)^4]\;.
\end{equation}
Comparing Eq.~(\ref{g0u3}) with Eq.~(\ref{g0uf}), it is evident that, since $Q \gg 1$, the presence of the cavity significantly amplifies the excitation rate of rotating atoms by a factor of 
$\frac{Q c^3}{V (2 \Omega -\omega _0)^{3}}$.

When $\Omega=\omega_{0}$ and $\Omega=2\omega_{0}/3$, the peak of the excitation rate at $\omega_{c}=2\Omega-\omega_{0}$ coincides with the peak of the emission rate at $\omega_{c}=\omega_{0}$ and $\omega_{c}=\omega_{0}-\Omega$, respectively. In this scenario, the emission rate (given by 
Eqs.~(6) and (7) 
in this Letter) is always larger than the excitation rate (given by Eq.~(\ref{g0u3})).

When $\Omega$ is not close to $\Omega=\omega_{0}$ and $\Omega=2\omega_{0}/3$ such that the peaks of the emission and excitation rates overlap, the emission rate for rotating atoms inside a cavity with the normal mode frequency tuned to $\omega_c=2\Omega-\omega_0$ is given by 
\begin{align}\label{g0d31}
\Gamma_{\downarrow}   \approx & \frac{2 \Omega -\omega _0}{6 \epsilon_{0} \hbar V  Q}\bigg\{\frac{2 \left[4 \Omega ^3+\omega _0 \left(\Omega ^2-8 \omega_{0} \Omega+4 \omega _0^2\right)\right] }{(\Omega-2\omega_0)^2 (3\Omega-2\omega_0)^2 
}\nonumber\\
&\times\left(d_{\rho }^2+d_{\phi }^2\right)+\frac{\omega _0 }{2 \left(\Omega -\omega _0\right){}^2}d_z^2\bigg\}+\mathcal{O}[\left({R\Omega}/{c}\right)^2]\;.
\end{align}
According to Eq.~(\ref{g0u3}) and Eq.~(\ref{g0d31}), it is possible for the excitation rate to surpass the emission rate if the parameters are chosen appropriately. 
For example, for isotropically polarizable atoms with a rotational angular velocity $\Omega=\frac{4\omega_0}{5}$, the ratio of the excitation rate to the emission rate is approximately $\frac{9 Q^2 R^2 \omega _0^2}{875 c^2}$. Consequently, the excitation rate can surpass the emission rate if the quality factor is sufficiently large such that $Q\gtrsim 10\frac{c}{R\omega_0}$. 

\begin{figure}[!b]
\centering
\subfigure[\;$\Omega >\omega_{0} $]{\includegraphics[scale=0.15]{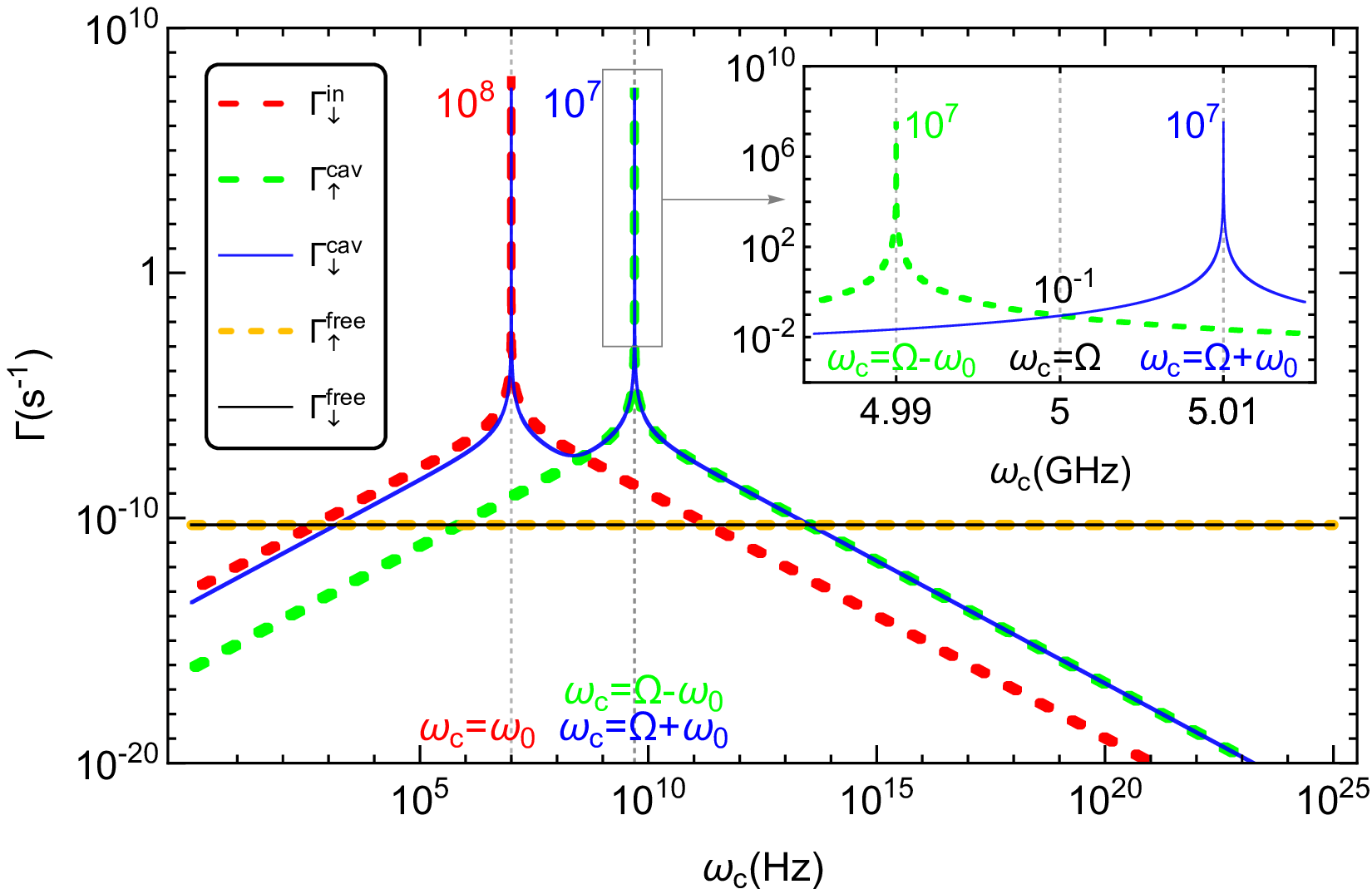}}
\label{figa}\\
\subfigure[ \;$\frac{\omega_{0}}{2}<\Omega <\omega_{0} $]{\includegraphics[scale=0.62]{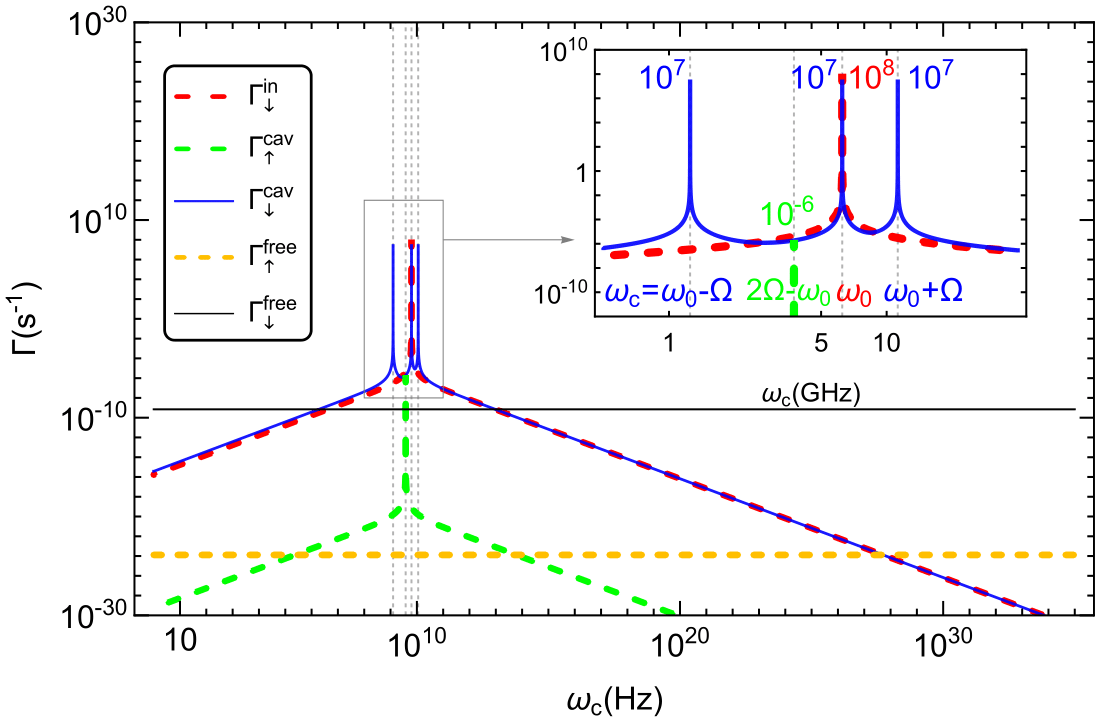}}\label{figb}
\caption{Emission rate $\Gamma_{\downarrow}$ and excitation rate $\Gamma_{\uparrow}$ as functions of the cavity's normal mode frequency $\omega_{c}$. 
The superscripts ``in", ``cav" and ``free" correspond to inertial atoms in the cavity, centripetally accelerated atoms in the cavity, and centripetally accelerated atoms in free space, respectively. The insets in panels (a) and (b)  show transition rates near the resonance conditions $\omega_c \approx \Omega$ and $\omega_c \approx 2\Omega - \omega_0$, respectively. 
The calculations assume isotropic atomic polarization and use experimentally feasible parameters from Ref.~\cite{Lochan20}: $d = 10^{-29} \, \text{Cm}$, $V=10^{-14} \, \text{m}^3$, $Q = 10^7$, $R = 50 \, \text{nm}$, $ \Omega=5\, \text{GHz}$, with (a) $\omega_{0} = 10 \, \text{MHz}$ and (b) $\omega_{0} = 6.25 \, \text{GHz} $. 
Note that in panel (a), the emission and excitation rates in free space do not overlap, satisfying  
$\Gamma_{\downarrow}^{\rm free}>\Gamma_{\uparrow}^{\rm free}$.} 
\label{Fig}
\end{figure}

\section{III. Numerical results of the transition rates}

Fig.~\ref{Fig} presents the emission and excitation rates of rotating atoms inside a cavity as functions of the cavity's normal mode frequency over a broad range,   alongside the corresponding rates in free space for comparison. 
The numerical results reveal prominent peaks at specific cavity frequencies, where both the emission and excitation rates of rotating atoms are significantly enhanced relative to those in free space, consistent with the analytical predictions. In particular, the excitation rate inside the cavity can be enhanced by 18 orders of magnitude compared with that in free space. In contrast, when the cavity's normal mode frequency is tuned far from these resonance conditions, both emission and excitation rates are substantially suppressed. This demonstrates the cavity's crucial role in not only enhancing but also inhibiting atomic transitions, offering a versatile mechanism for precise control over atomic dynamics.

\begin{figure*}[!t]
	\centering
	\subfigure[\;Excitation and emission rates as functions of the rotational angular frequency]{\includegraphics[scale=0.11]{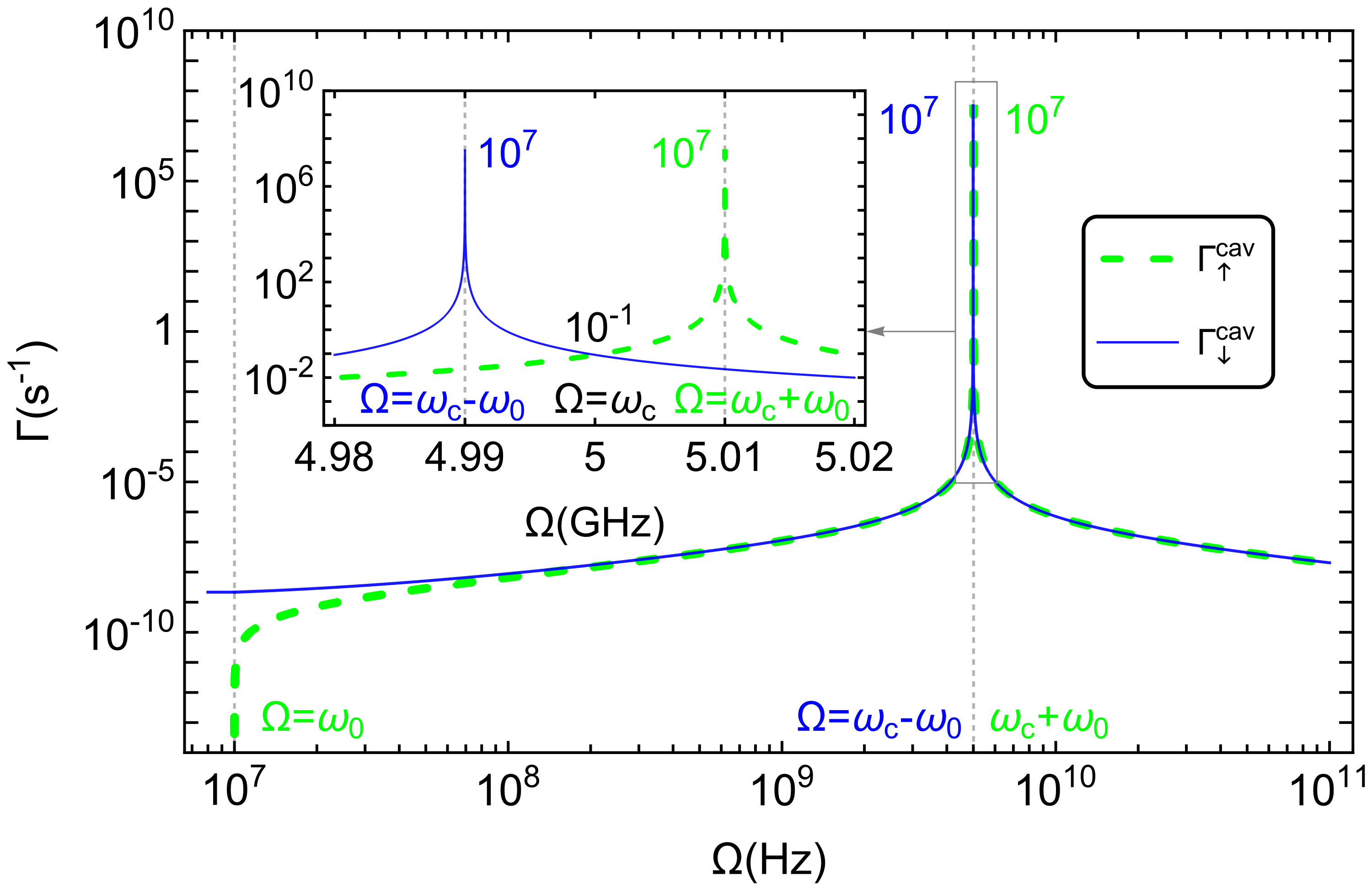}}
\label{Plotup}
	\subfigure[ \;Ratio of excitation to emission rates as a function of rotational angular frequency]{\includegraphics[scale=0.685]{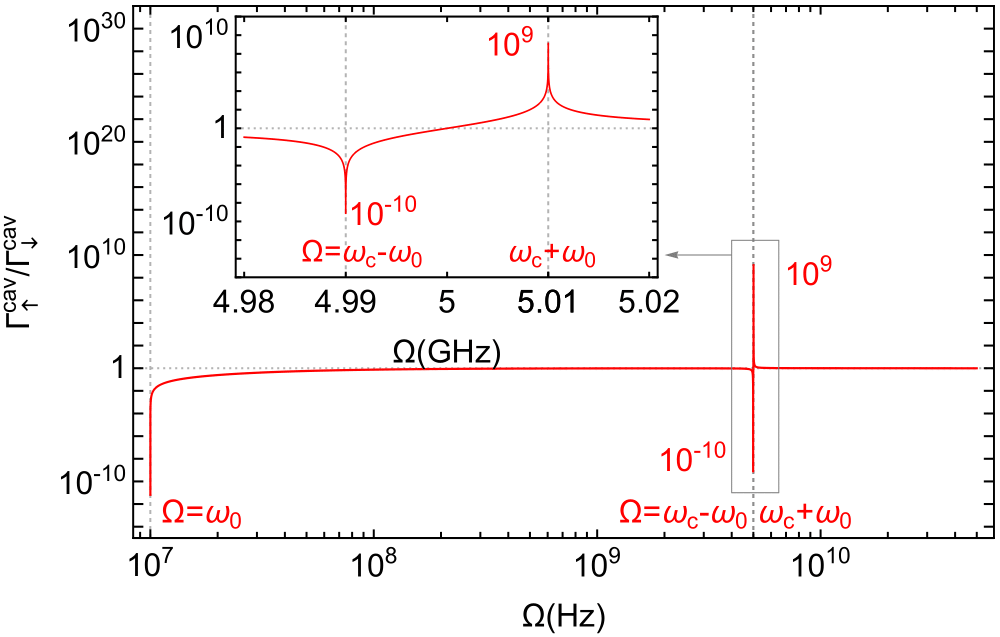}}\label{Plotuptodownrate}
\caption{Transition rates and the ratio of excitation to emission rates of a centripetally accelerated atom in a cavity as functions of the rotational angular velocity. Calculations assume isotropic atomic polarization and experimentally feasible parameters from Ref.~\cite{Lochan20}: $d = 10^{-29} \, \mathrm{Cm}$, $V = 10^{-14} \, \mathrm{m}^3$, $Q = 10^7$, $R = 50 \, \mathrm{nm}$, $\omega_0 = 10 \, \mathrm{MHz}$, and $\omega_c=5\, \mathrm{GHz}$. 
Note that here we consider only the regime where the rotational angular velocity exceeds the atomic transition frequency.}
	\label{Figomega}
\end{figure*}

Fig.~\ref{Figomega} (a) presents the excitation and emission rates of a centripetally accelerated atom in a cavity as functions of the rotational angular frequency. With the cavity properties fixed, the excitation rate increases with the rotational angular velocity and reaches a maximum of $10^7$ at $\Omega = \omega_c + \omega_0$. Similarly, the emission rate increases with the rotational angular frequency and peaks at $\Omega = \omega_c - \omega_0$, also with a maximum value of $10^7$. 
Fig.~\ref{Figomega}~(b) further presents the ratio of the excitation rate to the emission rate as a function of the rotational angular frequency.  
Once the rotational angular velocity exceeds the atomic transition frequency, the ratio increases rapidly and then approaches unity, indicating that the excitation and emission rates become nearly equal. When the frequency reaches $\Omega = \omega_c - \omega_0$, however, the ratio drops abruptly before rising again to a sharp peak at $\Omega = \omega_c + \omega_0$. Beyond this point, the ratio decreases sharply once more and asymptotically approaches unity. This behavior highlights both the intrinsically nonthermal nature of the circular Unruh effect and the cavity’s role in amplifying or suppressing the transition rates.

\section{IV. Transition rates of centripetally accelerated atoms in a cavity at finite temperature}

\begin{figure*}[!ht]
	\centering
	\subfigure[\;Emission rates: Rotating vs. inertial atoms inside a cavity immersed in a thermal bath]{\includegraphics[scale=0.62]{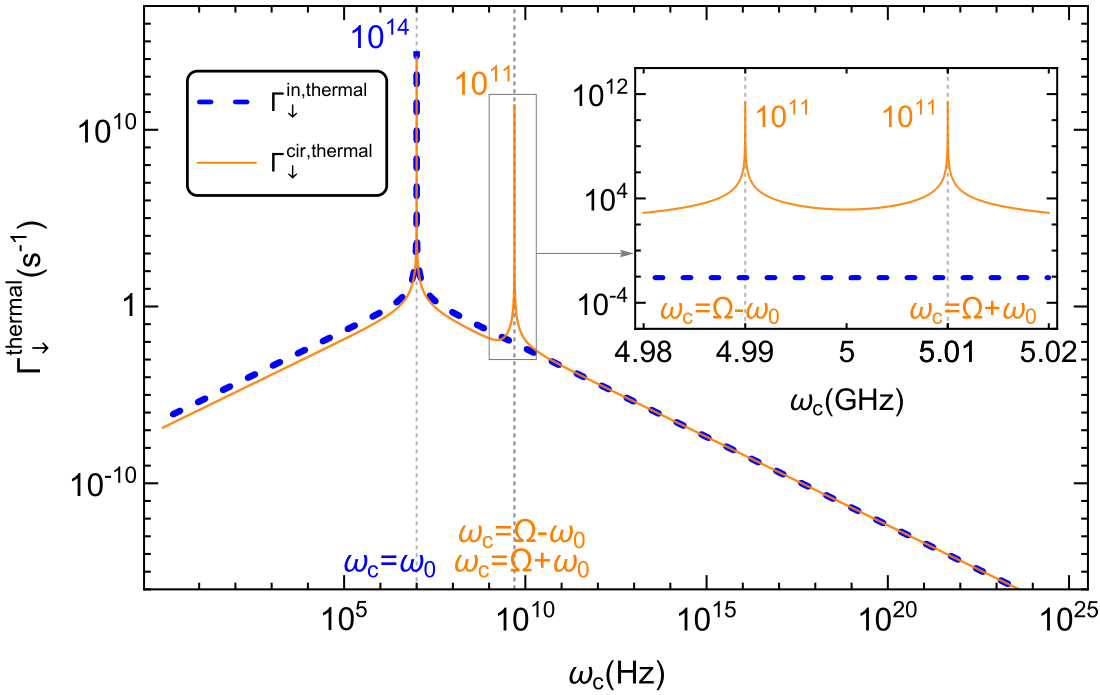}}
\label{figthermaldown}
	\subfigure[ \;Excitation rates: Rotating vs. inertial atoms inside a cavity immersed in a thermal bath]{\includegraphics[scale=0.62]{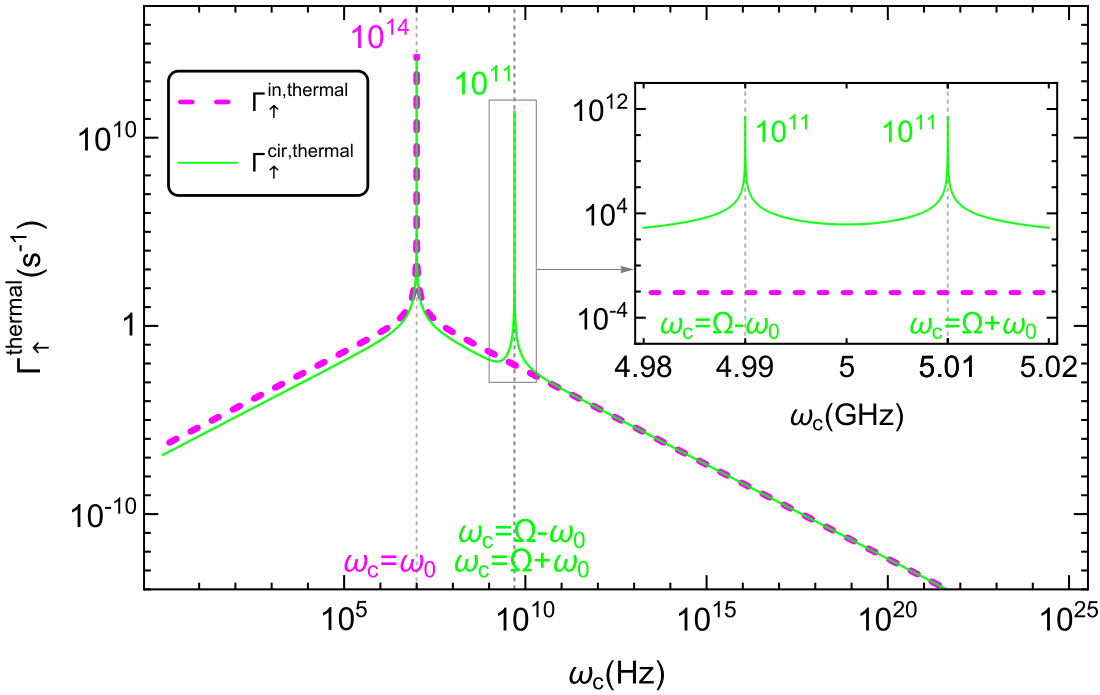}}\label{figthermalup}
\caption{Comparison between (a) the emission rate $\Gamma_\downarrow$ and (b) the excitation rate $\Gamma_\uparrow$ for rotating atoms and inertial ones inside a cavity immersed in a thermal bath. The superscripts ``in, thermal'' and ``cir, thermal'' denote inertial atoms and centripetally accelerated atoms inside the cavity, respectively. Insets in panels (a) and (b) show the transition rates in the vicinity of the resonance condition $\omega_c \approx \Omega$. Calculations assume isotropic atomic polarization and experimentally feasible parameters: $d = 10^{-29} \, \mathrm{Cm}$, $V = 10^{-14} \, \mathrm{m}^3$, $Q = 10^7$, $R = 50 \, \mathrm{nm}$, $\Omega = 5 \, \mathrm{GHz}$, $\omega_0 = 10 \, \mathrm{MHz}$, and $T = 300\, \mathrm{K}$.}
	\label{Figthermal}
\end{figure*}

Previously, we have assumed that the atoms are coupled with fluctuating electromagnetic fields in an ideal vacuum. However, in practical experiments, the environment cannot be a perfect vacuum but a thermal bath at finite temperature. In this case, the correlation functions in vacuum Eqs.~\eqref{correlationfunctionEE}-\eqref{correlationfunctionBE} should be replaced with the following ones representing those in a thermal bath at a temperature $T$,
\begin{widetext}
\vspace{-10pt}
\begin{align}
\left\langle \beta\left|E_{l}(t,\mathbf{x}) E_{p}\left(t',\mathbf{x}'\right)\right| \beta\right\rangle=&\frac{\hbar}{8 \pi \epsilon_{0} V}\int_{0}^{2 \pi} d\varphi \int_{0}^{\pi} \sin\theta d\theta \int_{0}^{\infty} d\omega_{k} \rho(\omega_k) \frac{\omega_k}{2} \left(\delta_{lp}-\frac{k_{l}k_{p}}{\boldsymbol{k}^{2}}\right) \nn\\
&\times \left[\left(1+\frac{1}{e^{\frac{\omega_{k} \hbar}{k_{B}T}}-1}\right)e^{-i( \omega_{k} t_{-}-\boldsymbol{k} \cdot \boldsymbol{R})}+\frac{1}{e^{\frac{\omega_{k} \hbar}{k_{B}T}}-1}e^{i( \omega_{k} t_{-}-\boldsymbol{k} \cdot \boldsymbol{R})}\right]\label{correlationfunctionEE}\;,\\
\left\langle \beta\left|B_{l}(t,\mathbf{x}) B_{p}\left(t',\mathbf{x}'\right)\right| \beta\right\rangle=&\frac{\hbar}{8 \pi\epsilon_{0} V}\int_{0}^{2 \pi} d\varphi \int_{0}^{\pi} \sin\theta d\theta \int_{0}^{\infty} d\omega_{k} \rho(\omega_k) \frac{\omega_k}{2c^2} \left(\delta_{lp}-\frac{k_{l}k_{p}}{\boldsymbol{k}^{2}}\right)  \nn\\
&\times \left[\left(1+\frac{1}{e^{\frac{\omega_{k} \hbar}{k_{B}T}}-1}\right)e^{-i( \omega_{k} t_{-}-\boldsymbol{k} \cdot \boldsymbol{R})}+\frac{1}{e^{\frac{\omega_{k} \hbar}{k_{B}T}}-1}e^{i( \omega_{k} t_{-}-\boldsymbol{k} \cdot \boldsymbol{R})}\right]\label{correlationfunctionBB}\;,\\
\left\langle \beta\left|E_{l}(t,\mathbf{x}) B_{p}\left(t',\mathbf{x}'\right)\right| \beta \right\rangle=&\frac{\hbar}{8 \pi \epsilon_{0} V}\int_{0}^{2 \pi} d\varphi \int_{0}^{\pi} \sin\theta d\theta\int_{0}^{\infty} d\omega_{k} \rho(\omega_k) \frac{\omega_k}{2c} \epsilon_{lpq} \frac{k_{q}}{\lvert \boldsymbol{k}\rvert} \nn\\
&\times \left[\left(1+\frac{1}{e^{\frac{\omega_{k} \hbar}{k_{B}T}}-1}\right)e^{-i( \omega_{k} t_{-}-\boldsymbol{k} \cdot \boldsymbol{R})}+\frac{1}{e^{\frac{\omega_{k} \hbar}{k_{B}T}}-1}e^{i( \omega_{k} t_{-}-\boldsymbol{k} \cdot \boldsymbol{R})}\right]\label{correlationfunctionEB}\;,\\
\left\langle \beta\left|B_{l}(t,\mathbf{x}) E_{p}\left(t',\mathbf{x}'\right)\right| \beta \right\rangle=&\frac{\hbar}{8 \pi \epsilon_{0} V}\int_{0}^{2 \pi} d\varphi \int_{0}^{\pi} \sin\theta d\theta\int_{0}^{\infty} d\omega_{k} \rho(\omega_k) \frac{\omega_k}{2c} \left(-\epsilon_{lpq} \frac{k_{q}}{\lvert \boldsymbol{k}\rvert}\right)  \nn\\
&\times \left[\left(1+\frac{1}{e^{\frac{\omega_{k} \hbar}{k_{B}T}}-1}\right)e^{-i( \omega_{k} t_{-}-\boldsymbol{k} \cdot \boldsymbol{R})}+\frac{1}{e^{\frac{\omega_{k} \hbar}{k_{B}T}}-1}e^{i( \omega_{k} t_{-}-\boldsymbol{k} \cdot \boldsymbol{R})}\right]\label{correlationfunctionBE}\;,
\end{align}
with $k_B$ denoting the Boltzmann constant. 
By substituting the explicit forms of the field correlation functions into Eq.~(2) in the Letter, the leading terms of the transition rates in the non-relativistic limit, i.e., when the linear velocity $v=R\Omega$ is much smaller than the speed of light $c$, can be calculated to be 
\begin{align}\label{thermaldown}
    \Gamma ^{\rm thermal}_{\downarrow}=&\frac{\pi }{6 V \epsilon_{0}  \hbar }\left\{\frac{2 \omega _0   e^{\frac{\omega _0 \hbar }{k_{B} T}}\rho\left(\omega _0\right)}{e^{\frac{\omega _0 \hbar }{k_{B} T}}-1}d_z^2+\left[\frac{\left(\omega _0+\Omega\right)  e^{\frac{ \left(\omega _0+\Omega \right) \hbar }{k_{B} T}}\rho\left(\omega _0+\Omega\right) }{e^{\frac{ \left(\omega _0+\Omega\right)\hbar }{k_{B} T}}-1} \right.\left.+\frac{ \left(\omega _0-\Omega\right) e^{\frac{(\omega _0 -\Omega ) \hbar }{k_{B} T}}\rho\left(|\omega _0-\Omega|\right) }{e^{\frac{(\omega _0 -\Omega ) \hbar }{k_{B} T}}-1} \right]\left(d_{\rho }^2+d_{\phi }^2\right) \right\}\nn\\
    &+\mathcal{O}\left[({R\Omega}/{c}\right)^2]\;,
\end{align}
%\begin{align}\label{thermaldown}
%    \Gamma ^{\rm thermal}_{\downarrow}=&\frac{\pi }{6 V \epsilon_{0}  \hbar }\left\{\frac{2 \omega _0   e^{\frac{\omega _0 \hbar }{k_{B} T}}\rho\left(\omega _0\right)}{e^{\frac{\omega _0 \hbar }{k_{B} T}}-1}d_z^2+\left[\frac{\left(\Omega +\omega _0\right)  e^{\frac{ \left(\Omega +\omega _0\right) \hbar }{k_{B} T}}\rho\left(\Omega +\omega _0\right) }{e^{\frac{ \left(\Omega +\omega _0\right)\hbar }{k_{B} T}}-1} \right.\left.+\frac{ \left(\Omega -\omega _0\right) \rho\left(|\Omega -\omega _0|\right) }{e^{\frac{(\Omega -\omega _0 ) \hbar }{k_{B} T}}-1} \right]\left(d_{\rho }^2+d_{\phi }^2\right) \right\}\nn\\
%    &+\mathcal{O}\left[({R\Omega}/{c}\right)^2]\;,
%\end{align}
\begin{align}\label{thermalup}
    \Gamma^{\rm thermal}_{\uparrow}=&\frac{\pi }{6 V \epsilon_{0}  \hbar }\left\{\frac{2 \omega _0\rho\left(\omega _0\right)}{e^{\frac{\omega _0 \hbar }{k_{B}T}}-1} d_z^2 +\left[\frac{\left(\omega _0+\Omega \right)  \rho\left(\omega _0+\Omega \right)}{e^{\frac{  \left(\omega _0+\Omega \right)\hbar}{k_{B}T}}-1}+\frac{\left(\omega _0 -\Omega\right)\rho\left(|\omega _0 -\Omega|\right) }{e^{\frac{(\omega _0 -\Omega)\hbar }{k_{B}T}}-1}\right]\left(d_{\rho }^2+d_{\phi }^2\right)\right\}\nn\\
    &+\mathcal{O}\left[({R\Omega}/{c}\right)^2]\;.
\end{align}
\end{widetext}
where $\mathcal{O}[x^n]$ denotes terms on the order of $x^n$ or higher that are negligibly small and therefore omitted. By setting $\Omega = 0$ in Eqs. (\ref{thermaldown})–(\ref{thermalup}), one directly obtains the inertial-case results.

We now present a numerical analysis of the transition rates for a centripetally accelerated atom inside a cavity immersed in a thermal bath at room temperature. These results are obtained using the same parameter set as in Fig.~\ref{Fig} (a), which corresponds to the vacuum case, but with the additional condition of $T = 300~\mathrm{K}$. The results are shown in Fig.~\ref{Figthermal}. 
When the normal mode frequency of the cavity is tuned to $\omega_c = \Omega - \omega_0$ and $\omega_c = \Omega + \omega_0$, the emission and excitation rates of the rotating atom reach $10^{11}\;\mathrm{s}^{-1}$, compared with only $10^{-3}\;\mathrm{s}^{-1}$ for an inertial atom, corresponding to an enhancement of $14$ orders of magnitude due to centripetal acceleration. 
Moreover, this substantial difference between the excitation rates of rotating and inertial atoms in a room-temperature cavity does not rely on fine tuning of the cavity resonance. Even when $\omega_c$ is slightly detuned from its optimal value, the excitation rate for the rotating case remains orders of magnitude larger than that for the inertial case. As shown in the insets of Fig.~\ref{Figthermal}  
(b), within the range $4.98-5.02\;\mathrm{GHz}$ the excitation rate for rotating atoms remains roughly five orders of magnitude higher, ensuring clear experimental distinguishability. 
These results demonstrate that rotation can strongly influence the transition rates of atoms inside the cavity even at room temperature, 
and that this enhancement is experimentally robust.

\end{document}